\documentclass{aa} 

\usepackage[version=4]{mhchem}
\usepackage[varg]{txfonts}
\usepackage{graphicx}
\usepackage{siunitx}
\usepackage{amssymb}                                
\usepackage[colorlinks,bookmarks]{hyperref}
\usepackage{color}
\usepackage[none]{hyphenat}
\definecolor{linkblue}{rgb}{0,0,0.8}
\definecolor{linkgreen}{rgb}{0,0.5,0}
\hypersetup{pdfpagemode=UseNone, pdfstartview=FitH, linkcolor=linkblue,citecolor=linkblue, urlcolor=linkblue}                    \usepackage{natbib}
\bibpunct{(}{)}{;}{a}{}{,}           

\begin{document}

% **********************************************
\def\be{\begin{equation}} 
\def\ee{\end{equation}} 
\def \dd{\mathrm{d}}
\def\erfc{\rm {erfc}} 
\def\bbf{\bf} 
\def\mpc{\,{\rm {Mpc}}} 
\def\mpch{\,h^{-1}{\rm {Mpc}}} 
\def\kms{\,{\rm {km\, s^{-1}}}} 
\def\vcir{{V_c}}  
\def\dt{{\Delta t}}  
\def\Gyr{{\,\rm Gyr}} 
\def\etal{{\it et al.~}}  
\def\Hm{${\rm {H^-}}\,\,$}  
\def\HH{${\rm {H_2}}\,\,$}  
\def\HHp{${\rm {H_2^+}}\,\,$} 
\def\sngg{SN$_{\gamma\gamma}$~}  
\def\fgg{f_{\gamma\gamma}} 
\def\sr{{\rm sr}} \def\hz{{\rm Hz}} \def\cm{${\rm cm}$} 
\def\nhi{{N_{\rm HI}}} \def\sec{{\rm s}} 
\def\sigmaba{\sigma_8/usr/local/lib/tex/inputs/latex/styles} 
\def\gsim{\lower.5ex\hbox{\gtsima}} 
\def\lsim{\lower.5ex\hbox{\ltsima}} \def\gtsima{$\; \buildrel > \over 
\sim \;$} \def\ltsima{$\; \buildrel < \over \sim \;$} \def\prosima{$\; 
\buildrel \propto \over \sim \;$} \def\gsim{\lower.5ex\hbox{\gtsima}} 
\def\lsim{\lower.5ex\hbox{\ltsima}} 
\def\simgt{\lower.5ex\hbox{\gtsima}} 
\def\simlt{\lower.5ex\hbox{\ltsima}} 
\def\simpr{\lower.5ex\hbox{\prosima}} \def\la{\lsim} \def\ga{\gsim} 
\def\ie{{\frenchspacing\it i.e. }} \def\eg{{\frenchspacing\it e.g. }} 
 \def\gtsima{$\; \buildrel > \over \sim \;$} 
\def\ltsima{$\; \buildrel < \over \sim \;$} 
\def\gsim{\lower.5ex\hbox{\gtsima}} 
\def\lsim{\lower.5ex\hbox{\ltsima}} 
\def\simgt{\lower.5ex\hbox{\gtsima}} 
\def\simlt{\lower.5ex\hbox{\ltsima}} 
\def\simpr{\lower.5ex\hbox{\prosima}} 
\def\la{\lsim} 
\def\ga{\gsim} 
\def\zcr{Z_{\rm cr}} 
\def\ekin{\mbox{\cal E}_{\rm kin}} 
\def\ekin{{\cal E}_{\rm kin}} 
\def\gg{\gamma\gamma} 
\def\fgg{f_{\gamma\gamma}} 
\def\Lya{Ly$\alpha$~} 
\def\sngg{SN-e$^{\pm}$~} 
\def\snggo{SN-e$^{\pm}$} 
\def\msun{\,{\rm \Msun}} 
\def\ie{{\frenchspacing\it i.e., }} 
\def\eg{{\frenchspacing\it e.g., }} 
\def\E3{{\cal E}_{\rm g}^{III}} 
\def\Eunit{\times10^{51} {\rm erg} \, \Msun^{-1}} 
\def\sEunit{10^{51} {\rm erg} \, \Msun^{-1}} 
\def\ozs{\Omega_Z^{sfh}} 
\def\ozo{\Omega_Z^{obs}} 
\def\Onot{\Omega_0} 
\def\Msun{\rm M_\odot}
\def\lsun{\rm L_\odot}
\def\Zsun{\rm Z_\odot}
\def\cmpc{\rm cMpc}
\def\kpc{\rm Kpc}
\def\Msun{\rm M_\odot}
\def\myr{\rm Myr}
\def\gyr{\rm Gyr }
\def\zsun{\rm Z_\odot}
\def\M*{M_*}
\def\mbh{M_{bh}}
\def\Z*{Z_*}
\def\L*{L_*}
\def\muv{\rm M_{UV}}
\def\EBV{E(B-V)}
\def\fws{f_*^w}
\def\luvs{L_*^{UV}}
\def\luvtot{L_{tot}^{UV}}
\def\fwb{f_{bh}^w }
\def\luvb{L_{bh}^{UV}}
\def\fs{f_*}
\def\fej{f_*^{\rm{ej}}}
\def\feff{f_*^{\rm{eff}}}
\def\highz{high-$z$\,}
\def \fesc{f_{\rm esc}}
\def \fescum {f_{\mathrm{esc}}^{\mathrm{cum}}}
\def \avfesc {\langle f_{\mathrm{esc}}\rangle}
\def\der{{\rm d}} 
\def\f{\frac}
\def\kev{\rm keV}
\def\mx{\,m_x} 
\def\K{\rm K}
\def \mges{M_*^{ge}}
\def \mgfs{M_*^{gf}}
\def \mgeb{M_{bh}^{ge}}
\def \mgfb{M_{bh}^{gf}}
\def\faccb{f_{bh}^{ac}}
\def\maccb{M_{bh}^{ac}}
\def\med{M_{ed}}
\def\fed{f_{ed}}
\def\mcritb{M_{bh}^{crit}}
\def\mdmsa{M_{dm}^{sa}}
\def\mgsa{M_{g}^{sa}}
\def\nho{n_{\mathrm{H}}^{0}}
\def\effesc{\fesc^{\rm eff}}
\newcommand{\pdc}[1]{\textcolor{teal}{[pdimp: #1\;]}}
\newcommand{\jb}[1]{\textcolor{blue}{[jb: #1\;]}}
\newcommand\code[1]{\textsc{\MakeLowercase{#1}}}
\newcommand{\quotes}[1]{``#1''}
\newcommand{\quotesing}[1]{`#1'}
\def\ie{{\frenchspacing\it i.e., }} 
\def\eg{{\frenchspacing\it e.g., }}
\newcommand{\footnoteref}[1]{\textsuperscript{\ref{#1}}}

\def \sixteenth{$16^{\rm{th}} \,$}
\def \eightyfourth{$84^{\rm{th}} \,$}
\def \fescuv{f_{\rm{esc}}^{\rm{UV}}}

\makeatletter
\@ifundefined{linenumbers}{}{%
  \let\linenumbers\relax
  \let\nolinenumbers\relax
}
\makeatother

\title{Effect of primordial black holes on the global 21 cm signal}

   \author{Atrideb Chatterjee\inst{1}\fnmsep\thanks{\email{a.chatterjee@rug.nl}}   }

   \institute{Kapteyn Astronomical Institute, University of Groningen, PO Box 800, 9700 AV Groningen, The Netherlands
 }

  \abstract
   {The 21 cm global signal, a treasure trove of information about the nature of the first luminous sources of the Universe, has traditionally been modelled assuming that these early sources were predominantly star-forming galaxies. However, recent observations by the James Webb Space Telescope (JWST) have revealed several active galactic nuclei (AGNs) as early as \( z \sim 10 - 10.4 \). In light of this, it is important to investigate the contribution of such AGNs to the 21 cm signal. Assuming that these AGNs are seeded by primordial black holes (PBHs) and employing an analytical PBH model, consistent with existing cosmological and astrophysical constraints, we show that these exotic objects can have a significant impact on the redshift evolution of the global signal.}


   \maketitle
% **********************************************

%%%%%%%%%%%%%%%%% BODY OF PAPER %%%%%%%%%%%%%%%%%%

\section{Introduction}

The sky-averaged 21 cm signal, also known as the global 21 cm signal, originates from the hyperfine transition of neutral hydrogen in the intergalactic medium (IGM). The redshift evolution of this signal carries information about the thermal and ionisation state of the IGM, which in turn is governed by the nature, timing, and intensity of the first astrophysical sources \citep[e.g.][]{pritchard2012, 2015MNRAS.449.4246G, 2017ApJ...848...23K, 2018MNRAS.477.3217G, 2019MNRAS.484..282G, 10.1088/2514-3433/ab4a73, 2021MNRAS.503.4551G, 2021MNRAS.507.2405C, Astraeus_I, Astraeus_VI}. As a result, the global 21 cm signal contains a wealth of information about the properties of these early sources. Most of the existing galaxy formation models assume these early sources to be of star-forming (SF) nature, with active galactic nuclei (AGNs) contributing significantly only at later times ($z \sim 5$) \citep{Onoue_2017, Dayal_20, Trebitsch_23, Dayal2025_Uncover}. 

Very recently, James Webb Space Telescope (JWST) observations have revealed that some of these early sources at \( z \sim 10 - 10.4 \) are in fact AGNs rather than SF galaxies \citep{bogdan2024, kovacs2024, napolitano2024}. Given that some of the properties of these AGNs (such as the elevated black hole-to-stellar mass ratios) are difficult to reconcile within the standard galaxy formation models \citep{matteri2025, Zhang_25, Nelander2025, dayal2025_PBH}, one of the alternative pathways to explain them is primordial black holes (PBHs) \citep{dayal2024_pbh, dayal2025_PBH, matteri2025, Nelander2025, Zhang_25}. If a fraction of PBHs, which originated during the time of inflation \citep{hawking1971, carr1974, carr2005, carr2020}, indeed seeded early galaxies, their accelerated formation and growth could have significantly influenced the thermal and ionisation state of the IGM at a very early epoch of the Universe when galaxy formation would not have yet been sufficiently underway — an effect unaccounted for in most of the 21 cm signal forecasts carries out so far \citep[e.g.][]{Barkana_loeb_2001, furlanetto2006, Chatterjee_23, Dhandha_25, Sims_25, Gessey_25}. Although, a number of studies \citep[e.g.][]{Tashiro13, EwallWice_20, Yang_21, Mittal_22, Nelander2025} indeed accounted for AGNs while modelling the 21 cm signal, their aim was to explain the non-standard feature of the 21 cm signal tentatively detected by the EDGES experiment \citep{Bowman2018}. This was later ruled out at a 95.3\% confidence level by the Shaped Antenna Measurement of the Background Radio Spectrum 3 (SARAS-3) experiment \citep{Saras_22}.

In this work, we investigated the imprints of PBH-seeded galaxies on the global 21 cm signal. We began with a simple semi-analytical model for SF galaxies, consistent with recent observations of the high-redshift Universe, and added the contribution of the PBH-seeded systems to determine the impact of these exotic objects on the 21 cm signal. While similar in spirit to \cite{Nelander2025}, our approach employs a more fundamental model for the PBH-seeded systems based on \cite{dayal2025_PBH}. Furthermore, the PBH model is consistent with the existing cosmological and astrophysical constraints. This question is timely, given such ongoing and planned experiments targeting the global 21 cm signal as SARAS-3 \citep{Saras_22}, the Large-Aperture Experiment to Detect the Dark Ages (LEDA; \citealt{Greenhill_2012}), SCI-HI \citep{Voytek_2014}, the Broadband Instrument for Global Hydrogen Reionisation Signal (BIGHORNS; \citealt{Sokolowski_2015}), the Radio Experiment for the Analysis of Cosmic Hydrogen (REACH; \citealt{Reach_22}), and the Cosmic Twilight Polarimeter (CTP; \citealt{Nahn_2018}).
Throughout this paper, we adopt a  Lambda cold dark
matter ($\Lambda$CDM) model with dark
energy, dark matter, and baryonic densities in units of the
critical density: $\Omega_{\Lambda} = 0.685$, $\Omega_m = 0.315$, and $\Omega_b = 0.049$, respectively. We use a Hubble constant $H_{0} = 100 h \, \rm{km\, s^{-1}\, Mpc^{-1}}$ with $h=0.67$, spectral index $n_s = 0.96$, and normalisation $\sigma_8 = 0.81$ \citep{Planck_20}.

The paper is organised as follows. In Sect. \ref{sec:theo_model}, we describe the theoretical model of the SF galaxies and the PBH-seeded systems. While Sect. \ref{sec:21cm_signal} presents the modelling of the 21 cm signal, Sect. \ref{sec:results} describes the effect of PBHs on the 21 cm signal. We summarise our results and conclusions in Sect. \ref{sec:discussion}. 

% *************************
\section{Theoretical model}
\label{sec:theo_model}

Our theoretical framework comprises two distinct components. The first accounts for early SF galaxies, whereas the second component models a population of PBHs that assemble their host galaxies around themselves.

% ********************   
\subsection{The semi-analytic model for star-forming galaxies}
As this work aims to determine the effect of PBH-seeded galaxies on the global 21 cm signal, any standard model of SF galaxies would suffice, so long as it does not violate any existing observational constraints on the high-z Universe (i.e. reionisation and UV luminosity function). Keeping this in mind, we used a very simple model for SF galaxies at $z \gtrsim 6$, as described below.
\begin{enumerate}
\item In this model, the star formation rate density (SFRD) at a redshift z is given by
\begin{equation}
    \rho_{\rm SFR} = \frac{1}{t_{\rm dyn}(z)}\frac{\Omega_b}{\Omega_m}\int_{M_{\min}(z)} dM_{h} \frac{dn}{dM_{h}}f_{\star}(M_{h}) M_{h}
\label{eq:rho_SFRD},
\end{equation}
with $M_h$ being the halo mass and $t_{\rm dyn}(z)$ the dynamical timescale at redshift z. The halo mass function, $\frac{dn}{dM_{h}}$, given by \cite{Tinker_2008} was implemented using the publicly available python package \texttt{colossus}\citep{hmf_code_18}. The factor $t_{\rm dyn} (z)$ appears in the denominator on the right-hand side of the equation, as we assume star formation in haloes to be spread over the entire $t_{\rm dyn} (z)$, following \cite{Chiu_and_Ostriker_2000}. Finally, $M_{\rm min}(z)$ is the minimum halo mass required for star formation. Under the simplified assumption that atomic cooling is the sole cooling channel, we determine $M_{\rm min}(z)$ corresponding to a virial temperature of $ 10^4$ K \citep{Barkana_loeb_2001} \footnote{We would like to point out that this simplified assumption is a reasonable approximation for the high redshift range adopted \citep{Barkana_loeb_2001, furlanetto2006} }. Following \cite{Donan_24}, we take the star formation efficiency to be
    \begin{equation}
        f_{\star} (M_h) = \frac{2 f_{0}}{(M_h/M_p)^{-\beta}+(M_h/M_p)^{\gamma}},
    \end{equation}
with $\{ f_{0} , M_{p} , \beta, \gamma) = \{0.16, 10^{11.7} , 0.9, 0.65\}$. This choice of star formation efficiency ensures that the UV luminosity function (UVLF) predicted from this model is consistent with the measurement from high-redshift JWST observations \citep{Donan_24}. We also note that this functional form is not physically motivated but is rather obtained from a data-driven point of view.

\item The number of photons produced per unit time at frequency $\nu$ at a redshift z was computed using
\begin{equation}
   \dot{n}_{\rm \nu} = \frac{dN_{\nu}}{dM} \times \rho_{\rm SFR},
\end{equation}
where $\frac{dN_{\nu}}{dM}$ is the number of photons produced at frequency $\nu$ per unit of stellar mass. We calculated this from \texttt{Starburst99} \citep{Starburst_99} assuming a standard Salpeter initial mass function (IMF) in the $1 - 100$ $M_{\odot}$ mass range with a metallicity of $0.05 M_{\odot}$.

\item The number of ionising photons intrinsically produced inside galaxies at a redshift z was computed as
\begin{equation}
    \dot{n}^{\rm ion}(z) = \int^{\infty}_{\nu_H} \dot{n}_{\rm \nu} (z) d\nu,
\end{equation}
where $\nu_{H}$ is the threshold frequency for hydrogen photoionisation.

\item The number of ionising photons escaping a galaxy that can reionise the IGM is given by $\dot{n}^{\rm SF}_{\rm ion, esc} = f_{\rm esc} \dot{n}^{\rm int}_{\rm ion}$. We fixed the value of the ionising escape fraction to $f_{\rm esc} = 0.2$, making sure that the model matches the existing reionisation constraints as shown in the Appendix-\ref{app:A}.

\item The effect of the astrophysically produced, i.e. `normal' AGNs, is ignored in this model. This is justified, given that the effect of these `normal' AGN dominates around $z \sim 5$ \citep{Dayal2025_Uncover}, whereas our redshift range of interest lies at a higher redshift, i.e. $z \sim 30-6$.

\end{enumerate}

% ***********************
\subsection{The formation and evolution of PBH-seeded galaxies}
In modelling the PBH-seeded galaxy population, we closely followed the analytic framework, \texttt{PHANES}, proposed by \cite{dayal2025_PBH} except for the change in the mass function of the PBH-seeded galaxies\footnote{We chose the $s=0$ scenario of the \texttt{PHANES} model, since the non-spinning model produces a better fit with the observations, as shown in Figs. 4 and 5 of \cite{dayal2025_PBH}}. We also rigorously verified that the model, even after changing the mass function, remained consistent with the existing cosmological and astrophysical constraints. Here, we briefly outline the salient (and relevant for a 21 cm signal) features and the changes we made to this model.
\begin{enumerate}

\item In \cite{dayal2025_PBH}, the authors used the simplified power-law mass function. In this work by contrast, we assumed the more popular lognormal function, as this is a reasonable approximation for a range of PBH mechanisms \citep{Clesse_15, Blinnikov_16}. Following \cite{Carr_75}, we took

\begin{equation}
    \frac{dN}{dM_{\rm PBH}} = \frac{\kappa}{M_{\rm PBH}^2\sqrt{2\pi\sigma^2}}\exp\left[ -\frac{1}{2}\left(\frac{ln(M_{\rm PBH}/M_c)}{\sigma}
    \right)^2\right],
\end{equation}
with $\kappa$ being the normalising constant, $M_c$ the characteristic mass of PBHs, and $\sigma$ the standard deviation of the distribution. The value of $\kappa$ was determined by matching the observed mass function value of $10^{-5.27} \rm{cMpc^{-3}}$ at $z \sim 10$ \citep{kovacs2024, bogdan2024}, assuming an average seed mass of $10^{3.65} \rm{M_{\odot}}$. Further, we assumed $\sigma = 0.7$ \citep{matteri2025} and took the characteristic mass, $M_c$, to be equal to the average seed mass, i.e. $M_c = 10^{3.65} \rm{M_{\odot}}$.

\item The bolometric luminosity of a PBH-seeded galaxy was calculated from
\begin{equation}
    L_{\rm bol} = \frac{\epsilon_rM^{\rm ac}_{\rm BH} c^2}{\Delta t } L_{\odot},
\end{equation}
where c is the speed of light, $\rm L_{\odot}$ is the solar luminosity, $\epsilon_r$ is the radiative efficiency, and $M^{\rm ac}_{\rm BH}$ is the accreted BH mass within one time step $\Delta t$. While $\epsilon_r$ was taken as $0.057$ \citep{dayal2025_PBH}, the accreted BH mass $M^{\rm ac}_{\rm BH}$ was calculated assuming an Eddington fraction $f_{\rm edd} =0.25$, following the procedure described in detail in \cite{dayal2024_pbh}.

\item  We calculated the photon production rate at a frequency ($\nu$) at redshift z from a PBH-seeded galaxy with
\begin{equation}
    \dot{n}^{\rm PBH}_{\rm \nu} =  \int dM_{\rm PBH} \frac{dN}{dM_{\rm PBH}} \frac{L_{\nu}(M_{\rm PBH})}{h_{p}\nu},
\end{equation}
where $L_{\nu}$ is the luminosity of individual galaxies at frequency $\nu$. For each galaxy, we first computed the B-band luminosity ($L_{B}$) from the bolometric luminosity, using the fitting formula from \cite{marconi2004}. We then assumed a spectral index of $-0.57$ $(-1.57)$ to compute Ly-$\alpha$ (ionising) photon production rate from $L_{B}$.

\item The escape fraction of the ionising photons is given by $f^{\rm PBH}_{\rm esc} = \exp(-\tau_{\rm PBH})$, where $\tau_{\rm PBH}$ is the dust optical depth determined by the metal enrichment and the Eddington fraction of the PBH-seeded galaxy, as described in detail in \cite{dayal2025_PBH} and \cite{Dayal2025_Uncover}. Therefore, the total number of ionising photons escaping into the IGM is given by $\dot{n}^{\rm PBH}_{\rm ion, esc} = f^{\rm PBH}_{\rm esc} \dot{n}^{\rm PBH}_{\rm ion, int}$, with $\dot{n}^{\rm PBH}_{\rm ion, int}$ being the total number of ionising photons produced inside the PBH-seeded system.

\item The X-ray luminosity in the $2-10$ keV range \citep{Furlanetto_06} for individual systems was calculated from their bolometric luminosity ($L_{\rm bol}$), using a correction factor $k_{\rm X} \equiv L_{\rm bol}/L_{\rm X}$ as described in \cite{duras2020}. The fitting function used to calculate $k_X$ is given by $k_{\rm X} = a\left[ 1 + \left(\frac{\log_{10}(L_{\rm bol}/L_{\odot}) }{b}\right)^c \right]$, where $a=10.96$, $b=11.93$, and $c=17.79$ are the values for the best-fit free parameters.
Once we calculated $L^{\rm PBH}_{\rm X}$ for the individual PBH-seeded system, we obtained the total X-ray emissivity at redshift z using
\begin{equation}
\label{eq: eX_PBH}
 \epsilon^{\rm PBH}_{X} = \int dM_{\rm PBH} \frac{dN}{d M_{\rm PBH}} \cdot L^{\rm PBH}_{X}.
\end{equation}

\end{enumerate}

\begin{figure*}
    \centering
    \includegraphics[width=\textwidth]{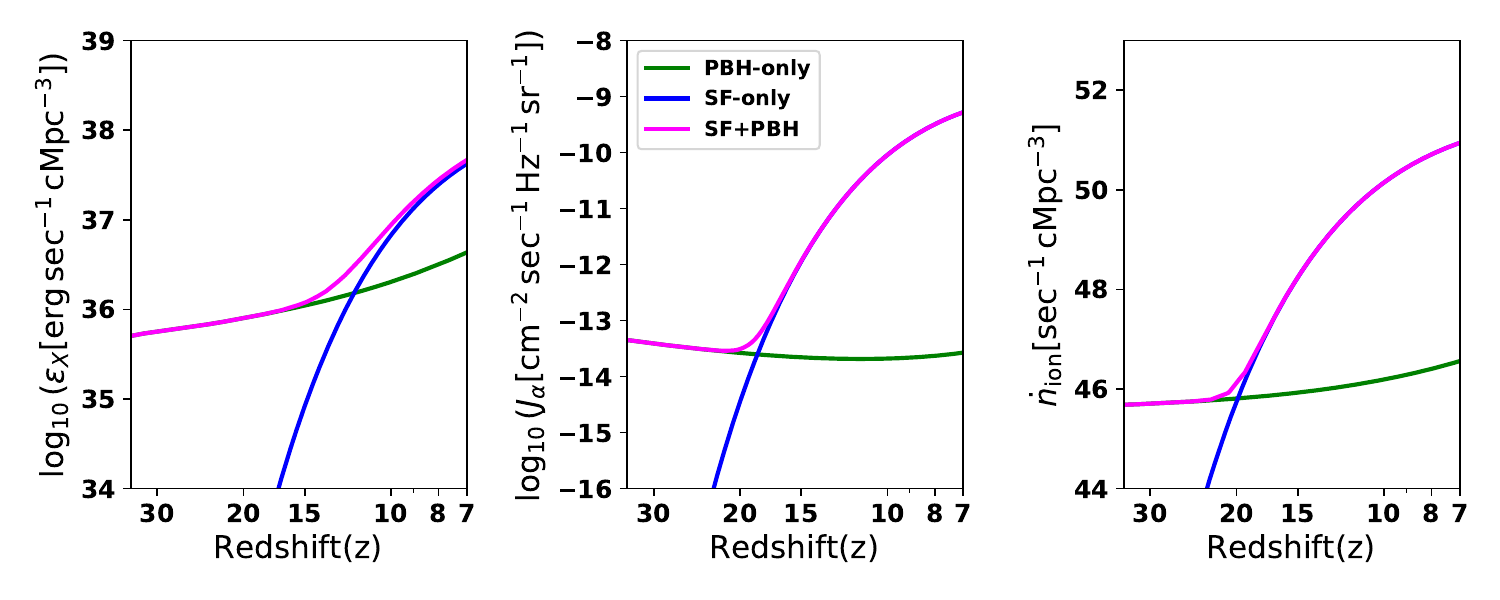}
    \caption{Redshift evolution of key quantities affecting the 21 cm signal. The green, blue, and magenta curves correspond to contributions from PBH-only, SF-only, and SF plus PBH galaxies, respectively. Leftmost panel: Intrinsic X-ray emissivity as a function of redshift. In this case, PBH-only dominates over SF-only for $30<z<15$. 
Middle panel: Evolution of the Ly-\(\alpha\) background flux. While dominated by PBH-seeded sources for the very early redshift around $ z = 30 - 20$, the SF-only galaxy completely dominates the redshift $z=20-7$. Rightmost panel: Ionising photon emissivity, which mimics the Ly-\(\alpha\) background trends.}
    \label{fig:all_quan}
\end{figure*}

% *****************
\section{The 21 cm global signal}
\label{sec:21cm_signal}

The sky-averaged 21 cm brightness temperature can be written as \citep{chatterjee_20}
\begin{equation}
    T_{b}(\nu) = 10.1 \, {\rm mK}\,\, \, x_{\rm HI}(z))\left( 1- \frac{T_{\gamma}(z)}{T_{S}(z)} \right) (1+z)^{1/2},
\end{equation}
where $x_{\rm HI}(z)$ is the neutral hydrogen fraction of the IGM at redshift z and $T_{\gamma}(z)$ is the background cosmic microwave background (CMB) temperature given by $T_{\gamma} = 2.73\,(1+z)\, {\rm K}$. Furthermore, the spin temperature ( $T_{S}$) of the hydrogen atom can be written as
\begin{equation}                            T_S^{-1}=\frac{T_{\gamma}^{-1}+x_{\alpha}T_{K}^{-1} + x_{c}T^{-1}_{K}}{1+ x_{c} + x_{\alpha}},
    \label{eq:tspin}
\end{equation}
where $x_{\alpha}$ is the Lyman-Alpha (Ly-$\alpha$) coupling coefficients, $x_{c}$ is the collisional coupling coefficient, and $T_{K}$ is the kinetic temperature of the IGM. Moreover, the collisional coupling coefficient, $x_c$, was determined using the standard formalism and the fitting functions provided in \cite{pritchard2012}. We describe all the other terms in more detail below.

\subsection{Kinetic temperature of the IGM}
The two main processes that determine the redshift evolution of the IGM kinetic temperature ($T_{K}$) consist of adiabatic cooling, due to the expansion of the Universe, and X-ray heating of the IGM. While adiabatic cooling was trivially computed following \cite{furlanetto2006}, computing X-ray heating is highly uncertain given our poor knowledge of the high-redshift Universe. Assuming that the relation between X-ray luminosity and the SFRD in the high-z Universe follows from the local Universe, we computed X-ray emissivity in the $2-10$ keV range \footnote{assuming a power-law index of $-1.5$, i.e. $L_{X} \propto \nu^{-1.5}$.} \citep{Furlanetto_06, mineo2012}:
\begin{equation}
    \frac{\epsilon_{X}^{\rm SF}}{\rm J \,sec^{-1} \, Mpc^3} = 3.4 \times 10^{33} \frac{\rho_{\rm SFR}}{\rm M_{\odot} \, yr^{-1} \, Mpc^3},
\end{equation}
where $\rho_{\rm SFR}$ is the SFRD obtained from the SF galaxies as described in Eq. \ref{eq:rho_SFRD}.

The total X-ray escaping the SF and PBH-seeded galaxies (Eq. \ref{eq: eX_PBH}) and heating the IGM is given by

\begin{equation}
\epsilon^{\rm SF+PBH}_X(z) = f_{\rm h}[f^{\rm SF}_{X, \rm esc}\epsilon_{X}^{\rm SF} (z) +f^{\rm PBH}_{X, \rm esc}\epsilon^{\rm PBH}_X(z)],
\label{eq:e_X_sfr}
\end{equation}
where $f_{h}$ is the fraction of the total X-ray that heats the IGM and is fixed at 0.2 \citep{furlanetto2006}. Moreover, $f^{\rm SF}_{\rm X,esc}$ and $f^{\rm PBH}_{\rm X,esc}$ represent the escape fractions of X-ray from the SF and PBH galaxies, respectively. We also note that the uncertainties in the $\epsilon_X-\rho_{\rm SFR}$ relationship, i.e. Eq.~\ref{eq:e_X_sfr}, potentially arising from the poorly constrained nature of the SF galaxies at high redshift, can be effectively absorbed into the parameter $f^{\rm SF}_{\rm X, esc}$ (as it acts as a multiplicative factor). In this work, $f^{\rm SF}_{\rm X,esc}$ was kept fixed at 1, and $f^{\rm PBH}_{\rm X,esc}$ was varied in the range $0.1-1.0$. We discussed the effect of varying $f^{\rm PBH}_{\rm X,esc}$ in detail in Sect. \ref{sec:results}.

\subsection{Ly-$\alpha$ coupling}
The Ly-$\alpha$ coupling coefficient $x_{\alpha}$ is written as
\begin{equation}
    x_{\alpha} = 1.81 \times 10^{11} \, (1+z)^{-1} S_{\alpha} \frac{J_{\alpha}}{\rm cm^{-2} sec^{-1} Hz^{-1} sr^{-1}}.
\end{equation}
Here, $S_{\alpha}$ is taken to be order unity \citep{Furlanetto_06} coming from a detailed analysis of atomic physics. 

Finally, the background Ly-$\alpha$ flux $J_{\alpha}$ in the presence of both SF and PBH-seeded galaxies is given by
\begin{equation}
    J_{\alpha}^{\rm SF+PBH}=\frac{c}{4 \pi}(1+z)^3 \int^{z_{\rm max}}_{z}\left[f^{\rm SF}_{\alpha}\dot{n}^{\rm SF}_{\nu'}(z')+ f^{\rm PBH}_{\alpha}\dot{n}^{\rm PBH}_{\nu'}(z')\right] \left|\frac{d t'}{d z'}\right| dz',
    \label{eq:Ly_alpha background}
\end{equation}

where $\dot{n}^{\rm SF}_{\nu'}(z)$ is the photon production rate at a frequency $\nu'$ at a redshift z from SF galaxies and $\dot{n}_{\rm PBH}$ is the contribution from the PBHs, as mentioned before. Furthermore, $f^{\rm SF}_{\alpha}$ and $f^{\rm PBH}_{\alpha}$ represent the escaping fraction of the Ly-$\alpha$ photons from the SF and PBH-seeded galaxies. For the sake of simplicity, we assumed both to be 1. Finally, the integration upper limit, $z_{\rm max}$, was calculated using \citep{chatterjee_20}
\begin{equation}
    1 + z_{\rm max} = \frac{\nu_H}{\nu_{\alpha}}(1+z),
\end{equation}
with $\nu_{\alpha}$ denoting the Ly-$\alpha$ frequency. The upper frequency cutoff ensures that continuum ionising photons absorbed by the neutral IGM are excluded, since they do not contribute to the production of Ly-$\alpha$ photons in the background.

\subsection{Reionisation}

The neutral hydrogen fraction, $x_{\rm HI}$, determined from the evolution of the volume filling factor for ionised hydrogen ($\rm Q_{\rm HII}$), in the presence of both PBH and SF galaxies, can be written as
\begin{equation}
\begin{split}
    \frac{d Q_{\rm HII}}{dt} = \frac{\dot{n}^{\rm PBH}_{\rm ion, esc} + \dot{n}^{\rm SF}_{\rm ion, esc}}{n_{\rm H, com}}+ Q_{\rm HII}\alpha_{B} \, {\cal C}\, n_{\rm H, com} (1+z)^3,
\end{split}
\end{equation}
where $ n_{\rm H, com}$ is the hydrogen comoving number density,  ${\cal C}$ is the clumping factor of the IGM, and $\alpha_{B}$ is the (case B) recombination rate coefficient. The functional form of the clumping factor C was taken as $1+43z^{-1.71}$ \citep{pawlik2009}. Finally, $\dot{n}^{\rm SF}_{\rm ion, esc}$ and $\dot{n}^{\rm PBH}_{\rm ion, esc}$ denote the rate of ionising photons escaping out to the IGM from the SF and PBH galaxies, respectively.

\begin{figure*}
    \centering
    \includegraphics[width=\textwidth]{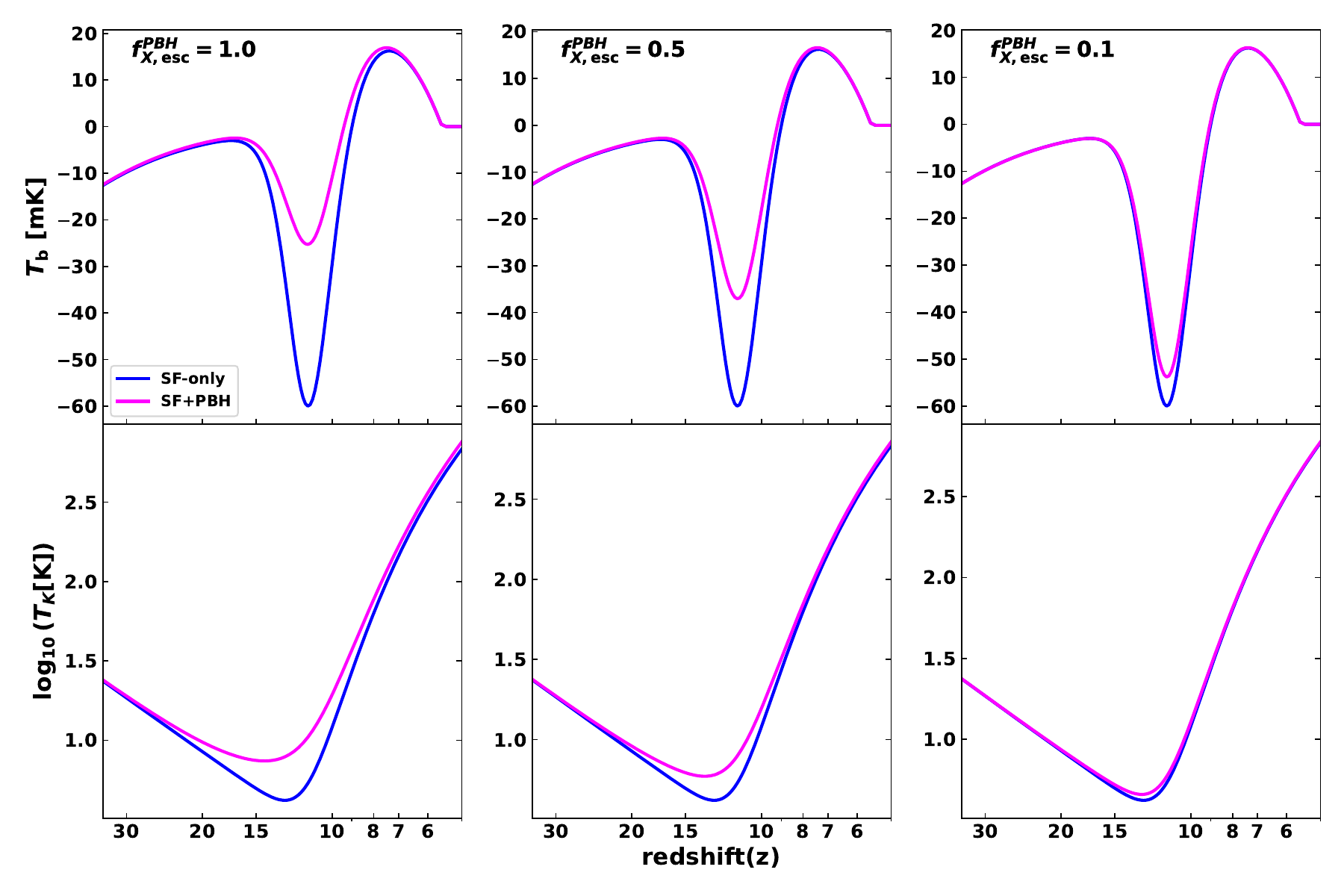}
    \caption{Top Panel: Redshift evolution of the global 21 cm signal for different values of $f^{\rm PBH}_{X, \rm esc}$ (keeping the escape fraction of X-rays from SF galaxies, $f^{\rm SF}_{\rm X,esc}$, fixed at 1.0), as indicated in the plots. The blue and magenta curves correspond to the SF-only and SF plus PBH scenarios, respectively. In the leftmost panel, with $f^{\rm PBH}_{X, \rm esc} = 1$, the SF-only case exhibits a much deeper absorption trough than the SF and PBH scenario. This difference arises because the additional X-ray emission from PBH-seeded galaxies produces excess heat, increasing the temperature of the IGM ($T_{k}$) as shown in the bottom panel of the figure.  As $f^{\rm PBH}_{X, \rm esc}$ decreases further (moving towards the right), the X-ray output from PBH-seeded galaxies and the corresponding $T_{k}$ diminish, leading to a progressively smaller difference between the two signals. We also note the difference in the signals around $25 \gtrsim z \gtrsim 20$. This can be attributed to the fact that SF and PBH together produce significantly more Ly-$\alpha$ photons compared to the SF-only model, as is also evident in panel (b) of \ref{fig:all_quan}.
    }
    \label{fig:T_b}
\end{figure*}

\section{Effect of PBHs on 21 cm signal}
\label{sec:results}

We begin by describing the redshift evolution of key quantities shown in different panels of Fig.~\ref{fig:all_quan}. The green, blue, and magenta curves correspond to the contributions from PBH-only, SF-only, and SF plus PBH galaxies, respectively.

In the left-most panel, we find the intrinsic X-ray emissivity at early redshifts ($z \sim 34-15$) to be dominated by PBH-seeded galaxies, with negligible contributions from the SF galaxies. It is only around $z \sim 15$ that the SF galaxy's contribution becomes non-negligible. This is expected, as the black holes in PBH-seeded AGNs accrete surrounding gas and produce copious amounts of X-ray photons before the onset of SF galaxies and therefore significantly outshine the emission from SF galaxies across the whole redshift range.

The middle and the rightmost panel show the Ly-$\alpha$ flux and ionising photon production rate. They are dominated by PBH-seeded systems in the very early redshift range $30 \gtrsim z \gtrsim 20$, the epoch when the SF galaxies are yet to form. However, once SF galaxies begin to emerge, they rapidly surpass the PBH-seeded systems in photon production around $z \sim 20$, indicating that stars are the primary sources of these photons.

To summarise, the X-ray photon budget is dominated by PBH-seeded galaxies in the redshift range $z \sim 34-15$, whereas the Ly-$\alpha$ and ionising photon contributions are dominated by PBHs only at very early times ($30 \gtrsim z \gtrsim 20$); at later times ($z \lesssim 20$), SF galaxies become the primary sources of Ly-$\alpha$ and ionising photons.

Next, we move to Fig.~\ref{fig:T_b}, where the top panel shows the global 21 cm signal for the SF-only and SF plus PBH models over the redshift range $34 \gtrsim z \lesssim 5$. The bottom panel shows the corresponding kinetic temperature of the IGM. In all the panels, the escape fraction of X-rays from SF galaxies, $f^{\rm SF}_{\rm X,esc}$, is fixed at 1, while the escape fraction from PBH-seeded galaxies, $f^{\rm PBH}_{\rm X,esc}$, decreases from $1.0$ to $0.1$ from left to right.

In the top-left panel, where $f^{\rm PBH}_{\rm X,esc} =1.0$ i.e. identical to that of the SF galaxies, the SF and PBH model produces a noticeably shallower absorption trough, with an amplitude of $\sim -25$ mK at $z \sim 12$. This contrasts with the SF-only model, which shows a deeper trough of amplitude $\sim -60$ mK at $z\sim 12$. This difference arises because, in the SF and PBH case, the X-ray contribution from PBH-seeded galaxies, dominating across the $30>z>15$ redshift range (as also shown in Fig. \ref{fig:all_quan}), enhances the temperature of the IGM ($T_{k}$), as shown in the corresponding bottom panel, with $T_{k}$ from SF and PBH $\sim 0.5$ dex higher than in the SF-only scenario.

As we move to the next panel on the right, where $f^{\rm PBH}_{X,\rm esc} = 0.5$ (i.e. the X-ray escape fraction of the PBHs is half that of the SF galaxies), the difference between the two signals begins to diminish. In the top-right panel, where $f^{\rm PBH}_{X,\rm esc}$ is one-tenth of that of SF galaxies, the amplitude of the absorption trough for the SF-only and SF plus PBH cases becomes nearly identical. A similar trend is observed in the bottom panel: the difference in the IGM temperature $T_k$ between the SF-only and SF and PBH scenarios decreases progressively as we move towards the right.

In each of the top panels, we also note that the signal from both SF and PBH in the redshift range $z=25-20$, near the first inflexion point, is slightly different from the SF-only scenario. This is attributed to the fact that the SF and PBH produces more Ly-$\alpha$ photons than the SF-only scenario, as also shown in the middle panel of Fig. \ref{fig:all_quan}. We further note that this difference in signal remains the same in all the panels as the Ly-$\alpha$ escape fraction is kept fixed for both models (i.e. $f^{\rm SF}_{\alpha} = f^{\rm PBH}_{\alpha} = 1$ across all the panels), confirming that this difference originates from Ly-$\alpha$ production independent of the X-ray emission in the models.

Next, we focus on the redshift range \( 10 \gtrsim z \gtrsim 7\) of the brightness temperature in the top panel of the figure. As is well known, the 21 cm signal in this regime is primarily determined by the neutral hydrogen fraction of the IGM. Considering that the SF-only model is already consistent with existing reionisation constraints (as shown in Appendix-\ref{app:A}) and that the contribution of PBHs \footnote{The contribution of the astrophysical black holes are shown to be negligible in \cite{Dayal2025_Uncover}.} to the ionising photon budget is negligible (as shown in panel (c) of Fig. \ref{fig:all_quan}), all models predict a similar neutral hydrogen fraction within this redshift range. Consequently, the resulting 21 cm signals from all models show similar behaviour.

\section{Conclusions and discussion}
\label{sec:discussion}
In this work, we investigated the effects of including PBH-seeded and SF galaxies on the global 21 cm signal. Our main findings are as follows.

\begin{itemize}
    \item Assuming that the escape fraction of X-ray photons from PBH-seeded galaxies is the same as that of SF galaxies, i.e. $f^{\rm PBH}_{\rm X,esc} = f^{\rm Sf}_{\rm X,esc}$, the 21 cm signal is strongly affected by the enhanced X-ray heating produced by the PBH-seeded systems. This additional heating makes the absorption trough of the 21 cm signal significantly shallower compared to the SF-only scenario. As we decrease the value of $f^{\rm PBH}_{\rm X,esc}$ from $1.0$ to $0.1$, the difference between these signals diminishes.
    \item The number of Ly-$\alpha$ photons coming from the PBH-seeded galaxies dominates the Ly-$\alpha$ photon budget in the redshift range $z = 30-20$ compared to those produced from SF galaxies; they alter the signal in this redshift regime $25 \gtrsim z \gtrsim 20$.
    \item Finally, the number of ionising photons coming from the PBH-seeded galaxies is 2-3 orders of magnitude lower compared to those produced from SF galaxies in the redshift range $z = 10-7$. Therefore, the introduction of PBHs leaves the reionisation constraints unaffected.
\end{itemize}
 
We end our paper by highlighting some shortcomings. First of all, we did not include the radio emission that may come from these PBH-seeded systems. This is justified since recent studies \citep{Mazzolari_25, Mazzolari_26} suggest that the radio emission from these early AGNs is significantly weaker compared to that from local AGNs. Secondly, there is no rigorous justification for choosing the lognormal mass function for the PBH-seeded systems. We plan to investigate the effect of different mass functions of PBHs on the 21 cm signal in future work. Third, we varied the X-ray escape fraction for the PBHs but did not vary the Ly-$\alpha$ photon escape fraction; in principle, one should vary both these parameters simultaneously. In the future, we also plan to address the effect of simultaneously varying different free parameters that appear in PBH-seeded galaxies using an MCMC framework.

%%%%%%%%%%%%%%%%%%%%%%%%%%%%%%%%%%%%%%%%%%%%%%%%%%%%%%%%%%%%%%
\begin{acknowledgements}
The work of AC was supported by the European Union’s Horizon Europe research and innovation programme under the Marie Skłodowska-Curie Postdoctoral Fellowship HORIZON-MSCA-2023-PF-01, grant agreement No 101151693 (LUPCOS). AC gratefully acknowledges Pratika Dayal, Tirthankar Roy Choudhury, Koushiki, Barun Maity and Amrita Banerjee for their constructive feedback on the manuscript. AC acknowledges the use of CHATGPT for refining the text at the final stage of the manuscript.
\end{acknowledgements}

\bibliographystyle{aa} % style aa.bst
\bibliography{main}

%%%%%%%%%%%%%%%%%%%%%%%%%%%%%%%%%%%%%%%%%%%%%%%%%%%%%%%%%%%%%%%
\begin{appendix}
\section{Redshift evolution of $Q_{\rm HII}$ for different models}
\label{app:A}
\begin{figure}
    \centering
    \includegraphics[width=\linewidth]{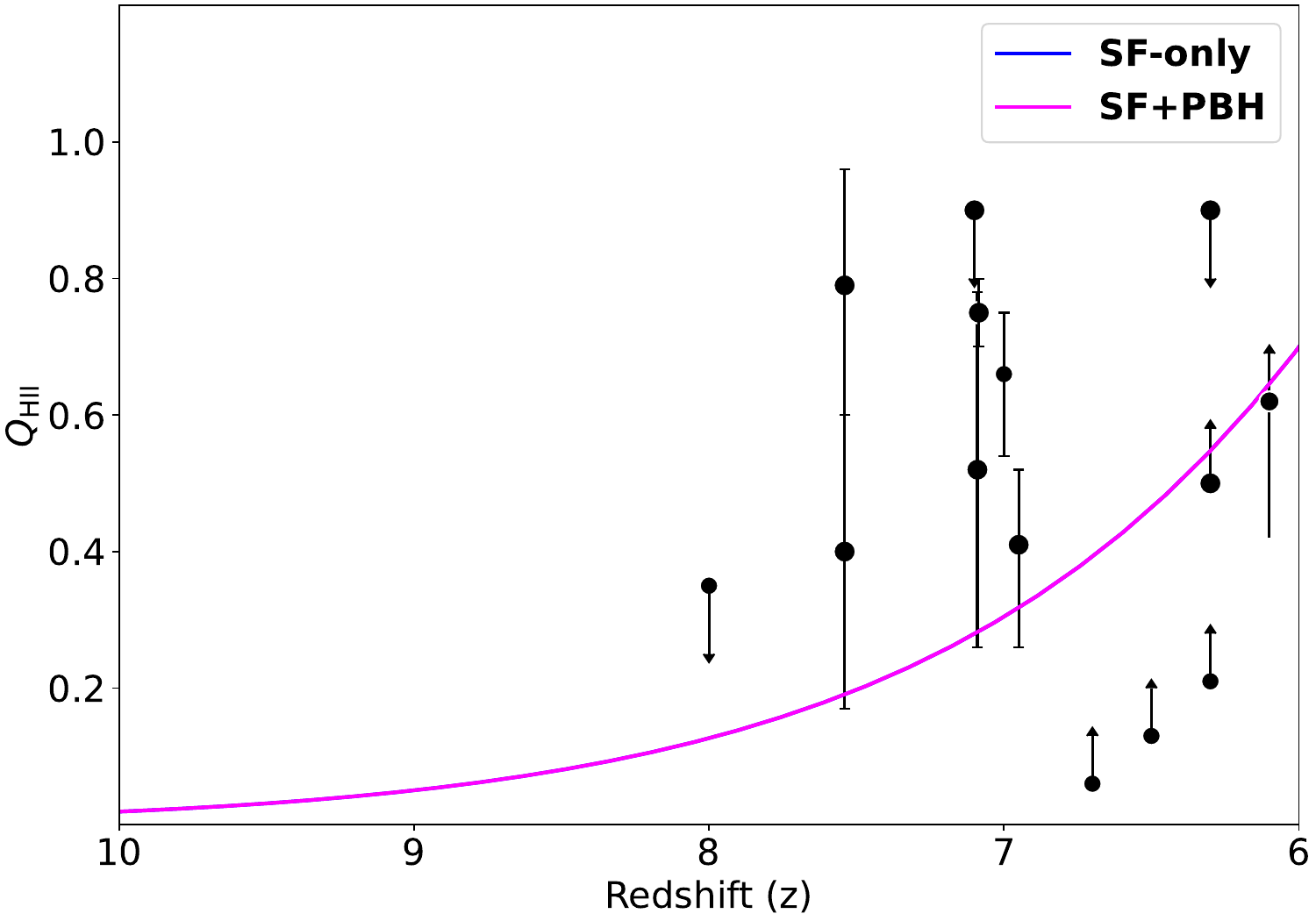}
    \caption{Redshift evolution of the volume filling factor of the HII region, $Q_{\rm HII}$. The data points are collected from different observations, as mentioned in the text.}
    \label{fig:QHII}
\end{figure}

Here we show the redshift evolution of the volume filling factor of the HII region, $Q_{\rm HII}$, coming from SF-Only and SF+PBH and how they compare with the existing observations related to reionisation. From Fig \ref{fig:QHII}, it is evident that the evolution of $Q_{\rm HII}$ coming from SF-Only and SF+PBH is practically identical, and they match quite well with the existing observational data points \citep{Totani_06, Chornock_13, Schroeder_13, Schenker_14, McGreer_15, Davies_18, Mason_18, Greig_19, Durovkov_20, Jin_23}. The  $Q_{\rm HII}$ from SF-Only and SF+PBH overlap because the PBH contribution to the ionising photon budget is negligible, as discussed in the main text.

\end{appendix}

\label{lastpage}
\end{document}